%% file: RaulRPrado_Auger.tex
\documentclass[epj,twocolumn]{webofc}
\usepackage[varg]{txfonts}   
\woctitle{ISVHECRI 2018}

\usepackage{xspace}
\usepackage{amsbsy}
\usepackage{rviewport}
\usepackage{comment}
\usepackage{cleveref}
\crefname{chapter}{Chap.}{Chaps.}
\crefname{appendix}{App.}{Apps.}
\crefname{section}{Sec.}{Secs.}
\crefname{paragraph}{Sec.}{Secs.}
\crefname{table}{Tab.}{Tabs.}
\crefname{figure}{Fig.}{Figs.}
\crefname{equation}{Eq.}{Eqs.}
\crefname{item}{item}{items}

\usepackage{lineno}

\input{TexAbbreviations}

\input{ThisAbbreviations}

\include{bib_aliases}

\begin{document}

\title{Tests of hadronic interactions with mea\-su\-re\-ments by Pierre Auger Ob\-ser\-va\-to\-ry}

\author{\firstname{Raul R.} \lastname{Prado}\inst{1,2,3}\fnsep\thanks{\email{raul.prado@desy.de}}
  for the Pierre Auger Collaboration\inst{4}\fnsep\thanks{Full author list: http://www.auger.org/archive/authors\_2018\_05.html} }
\institute{Deutsches Elektronen-Synchrotron (DESY), Platanenallee 6, D-15738 Zeuthen, Germany
\and
  IKP, Karlsruhe Institute of Technology (KIT), Postfach 3640, D-76021 Karlsruhe, Germany
\and
Instituto de F\'isica de S\~ao Carlos (IFSC/USP), Av. Trabalhador S\~ao-carlense 400, 13566-590, S\~ao Carlos, Brazil  	
\and	
	Av. San Martin Norte 304, 5613 Malarg\"{u}e, Argentina	
}

\abstract{
  The hybrid design of the Pierre Auger Observatory allows for the measurement of a number of properties of extensive air showers initiated by ultra-high energy cosmic rays. By comparing these measurements to predictions from air shower simulations, it is possible to both infer the cosmic ray's mass composition and test hadronic interactions beyond the energies reached by accelerators. In this paper, we will present a compilation of results of air shower measurements by Pierre Auger Observatory which are sensitive to the properties of hadronic interactions and can be used to constrain the hadronic interaction models. The inconsistencies found between the interpretation of different observables with regard to primary composition and between their measurements and simulations show that none of the currently used hadronic interaction models can provide a proper description of air showers and, in particular, of the muon production. 
}

\maketitle
 
\section{Introduction}
\label{sec:intro}

The Pierre Auger Observatory~\cite{\AugerPaper} 
is designed to measure ultra-high energy cosmic rays ($E>\E{17}$)
through the detection of extensive air showers (EAS). 
The two main parts of the 
observatory are the Fluorescence Detector (FD), composed of 
27 telescopes grouped in 4 sites and 
the Surface Detector (SD),
which consists of an array with 1660 water-Cherenkov 
detectors (WCDs) distributed over 3000 km$^2$.
The FD measurements allow the reconstruction of
the longitudinal profile of EAS, whereas the SD can 
measure the spatial distribution of particles on the ground
and their arrival times. The information obtained from both detectors
can be used to reconstruct a number of EAS observables. 
By comparing the measurements of these
observables to predictions from Monte Carlo simulations using
hadronic interaction models, one can either infer the 
cosmic-ray's mass composition or study 
the properties of hadronic interactions by testing the models. 
For each one of these goals, a different set of observables 
turns out to be more suitable.    
 
The most reliable observable for mass composition inference
is \xmax, the depth
of maximum energy deposit.
Lighter primaries penetrate deeper into the atmosphere than heavier ones,
producing on average showers with larger \xmax values. 
It is well known that the \xmax is mostly 
driven by the properties of only the first interaction, which
starts the dominant electromagnetic cascade producing most
of the electron and positron contribution to the energy 
deposit profile.  
Although the exact relation between \xmax, the primary energy, and 
mass depends on the properties of hadronic interactions,
the theoretical uncertainties on the predictions of \xmax by the most recent
hadronic models are relatively small when compared to other observables
(e.g. number of muons). The difference between the model predictions
of \xmaxmean is $\approx\pm20$ g/cm$^2$ (which translates into a 
difference in \lnamean of $\approx\pm 0.8$) 
and is constant over energy~\cite{PierogICRC2017}. 
Because of this, the cosmic-ray composition derived from the \xmax 
measurements is commonly used as a reference, 
allowing the hadronic models to be tested 
by comparing the composition interpretation from other observables
with that using \xmax. The \xmax measurements by Auger 
can be found in Ref.~\cite{\AugerXmaxPRDPaper,\AugerXmaxICRC2017Paper}. 

The muonic component of EASs is particularly interesting in the
context of testing hadronic interactions. 
Muons are mostly produced by the decay of charged mesons
when their energies are low enough such that their decay length
is comparable to the interaction length. Before reaching such low energies,
a number of generations of hadron-air interactions is required.
As a consequence, the muonic component is sensitive to the chain 
of hadronic interactions and, in particular, to their particle
production properties. It is well known
that the current hadronic models cannot provide a satisfactory 
description of the muon production in EAS. The most popular
manifestation of this fact is the so-called \emph{muon deficit problem},
i.e. the number of muons (\nmu) predicted by EAS simulations
is significantly smaller than that in measurements. 
This deficit has been observed by several
experiments~\cite{\HiResMIAMuonPaper,\IceTopMuonPaper,
	\AugerHASMuonPaper,\AugerTopDownPaper,
	Abbasi:2018fkz,\SugarMuonPaper}.
Therefore, experimentally accessing the muonic component 
of EAS, through the measurements of \nmu as well as further
observables sensitive to muons, is of extreme value in 
constraining hadronic models and 
possibly infer properties of hadronic interactions.

\begin{figure*}[!ht]
	\centering
	\includegraphics[width=0.49\textwidth]{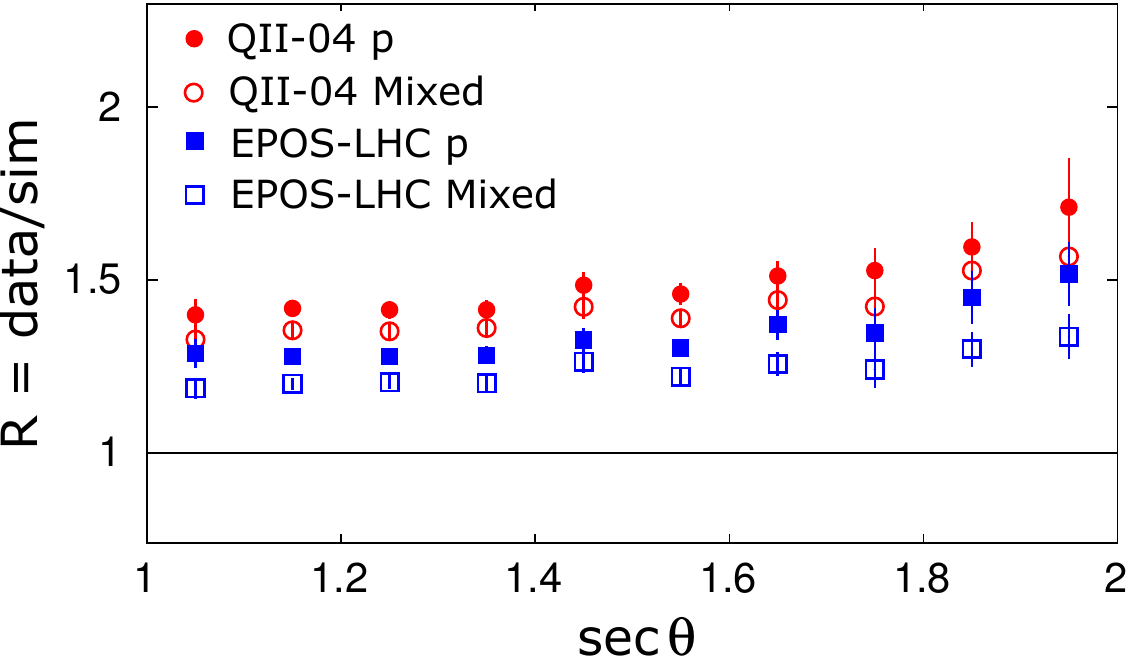}
	\includegraphics[width=0.43\textwidth]{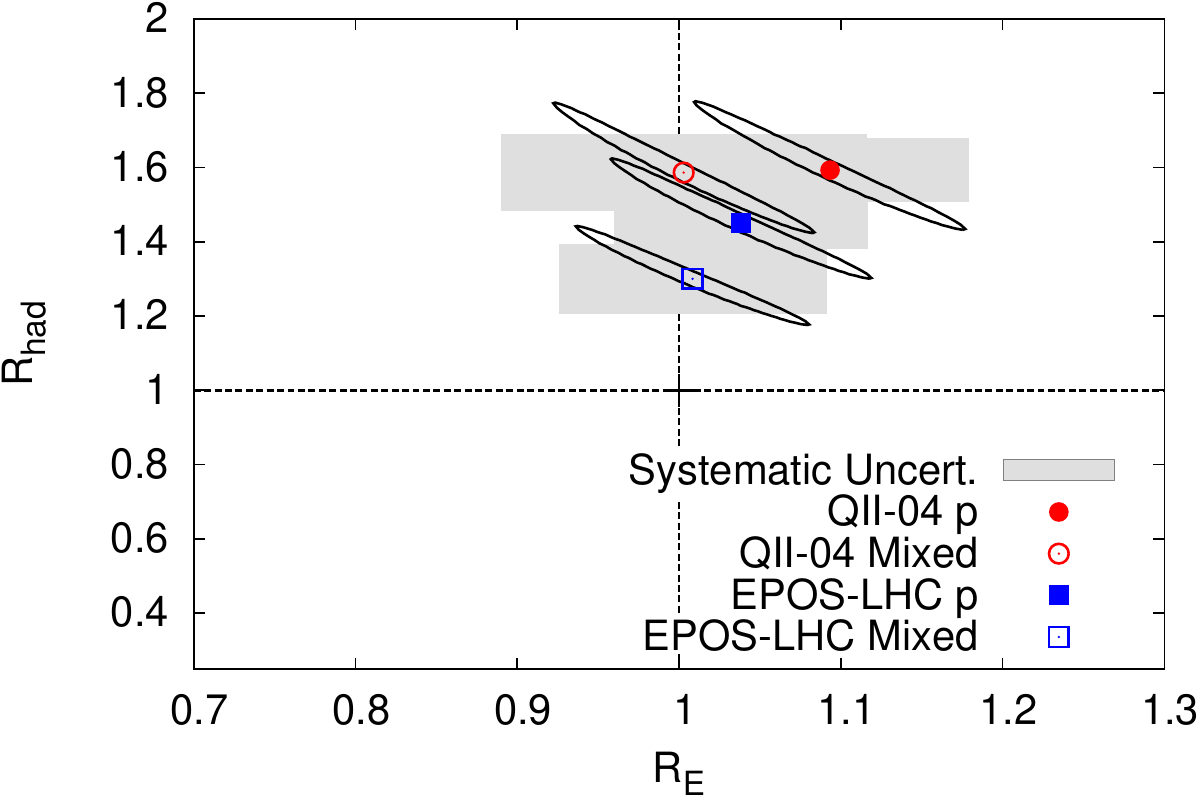}
	\vspace{-0.2cm}
	\caption{Results of the top-down analysis~\cite{\AugerTopDownPaper}. 
			See text for details.}
	\label{fig:topdown}
	\vspace{-0.4cm}
\end{figure*}

In this paper, we will present a selected compilation of 
measurements by the Pierre Auger Observatory which can be used 
to test hadronic interaction models. In~\cref{sec:topdown},
we present the results of the so-called Top-Down analysis, in which 
a set of hybrid events are compared in detail with simulations
in a way that it is possible to disentangle the contributions
from the electromagnetic and muonic components of EAS.  
In~\cref{sec:has,sec:mpd}, we present the results of two
analyses based on EAS observables which are purely muonic:
the number of muons measured in highly inclined events (\cref{sec:has})
and the muon production depth (\cref{sec:mpd}). In~\cref{sec:asym,sec:delta},
two analyses based on the parameter risetime (\rt) are presented:
the measurements of the azimuthal asymmetry of risetime(\cref{sec:asym})
and the so-called \ds-method (\cref{sec:delta}). 
Finally, in~\cref{sec:summary} we summarize the results. 

\section{Top-Down analysis}
\label{sec:topdown}

By using the experimental setup of Auger, 
the muon content of EAS can be directly measured 
in highly inclined events ($\theta > \dg{60}$) by using the atmospheric 
attenuation to eliminate the electromagnetic component (see~\cref{sec:has}).
However, it is not possible to isolate the muonic
from the electromagnetic component at the detector level for vertical events ($\theta < \dg{60}$).
To overcome this difficulty, an original analysis procedure was
developed aiming to evaluate the muon content in hybrid events
by means of a detailed comparison with Monte Carlo simulations. 
The full description of the analysis and the results 
can be found in Ref.~\cite{\AugerTopDownPaper}. 

For a set of 411 high quality hybrid events with 
$\e{18.8} < E / \text{eV}< \e{19.2}$, the first analysis step 
was to find simulated events in which the longitudinal profile
matches those measured by the FD. 
By construction, these simulated events have an 
\xmax and energy compatible with the corresponding measured event. 
The above procedure allows us to compare the SD signals
from measured and simulated events without having to account 
for differences in the longitudinal development of the showers.
The first step of this comparison can be seen 
in~\cref{fig:topdown} (left), where we show the average ratio $R$ 
between the shower size parameter \size obtained from data
and simulations as a function of the shower zenith angle $\sec(\theta)$. 
Apart from two different hadronic models, \EposLHCLong~\cite{\EposLHCPaper} 
and \QGSJetLong~\cite{\QGSJetPaper}, we also show two different composition
assumptions: proton only and a mixed composition scenario obtained
from the interpretation of the \xmax measurements by Auger.
It can be seen that for any model and composition scenario, the ratio $R$
is greater than $\approx1.2$, which shows that the ground signal 
of measured events is at least 20\% larger than the simulated ones with
the same longitudinal development. Furthermore, an evolution of
$R$ with zenith angle is observed, which will be important for the next step of the analysis.

In a second step, the particle distributions in simulated events were
rescaled in a way to make the data and simulations compatible in terms
of the shower size \size. Given that the different components of the shower
evolve differently with zenith angle and a large zenith angle range is
covered by the present analysis, it becomes possible to separately rescale
the electromagnetic component, represented by $R_E$, and the hadronic component,
represented by $R_\text{had}$. While $R_E$ is directly related to the 
shower energy scale, $R_\text{had}$ is responsible for scaling the 
muon content, meaning that by increasing $R_\text{had}$ the \nmu in the shower
would also increase by the same factor.  
A model derived from simulations was
used to perform the rescaling and the results 
can be seen in~\cref{fig:topdown} (right).  
To make the ground signal of simulations compatible with
those observed in the data, an increase from 30\% to 70\%
of $R_\text{had}$ is required, 
considering all the possible combinations of hadronic
models and composition assumptions.
On the other hand, it is observed that the energy scale factor
is never required to increase by more than 10\% and the case
without rescaling ($R_E$) is covered by the systematic
uncertainties in all cases. 

The results described above represent a manifestation of the
muon deficit problem. The best case scenario of the mixed 
composition assumption with \EposLHCLong as hadronic model
still requires an increase of 30\% on the \nmu predicted
by simulations at energies $\approx\E{19}$. For the first time, 
the muon deficit was evaluated by disentangling the contribution
of the muonic component from the energy scale.  

\begin{figure*}[!ht]
	\centering
	\includegraphics[rviewport=0.01 0 0.97 1, width=0.405\textwidth]{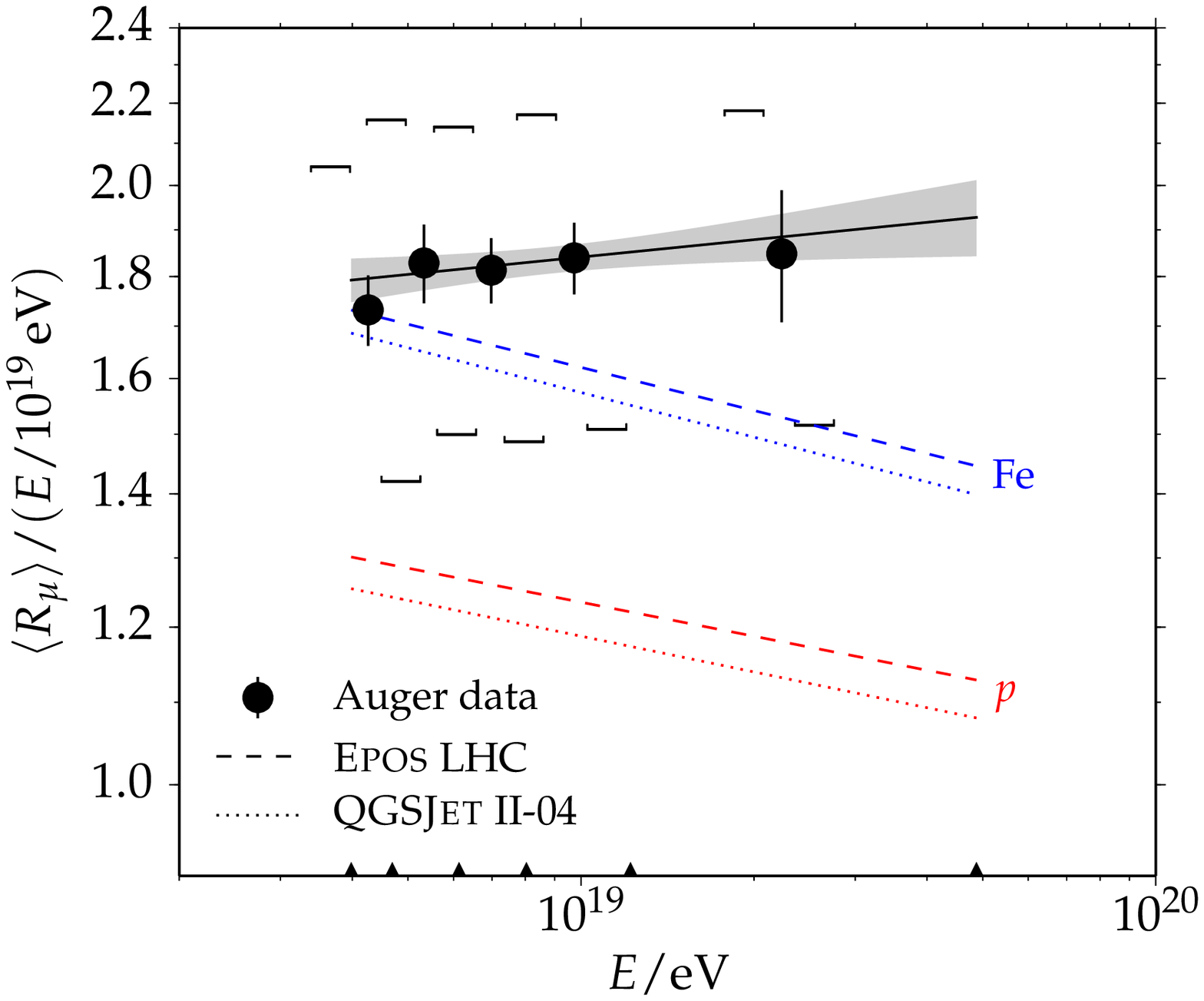}
	\hspace{1cm}
	\includegraphics[rviewport=0.07 0 0.95 1, width=0.39\textwidth]{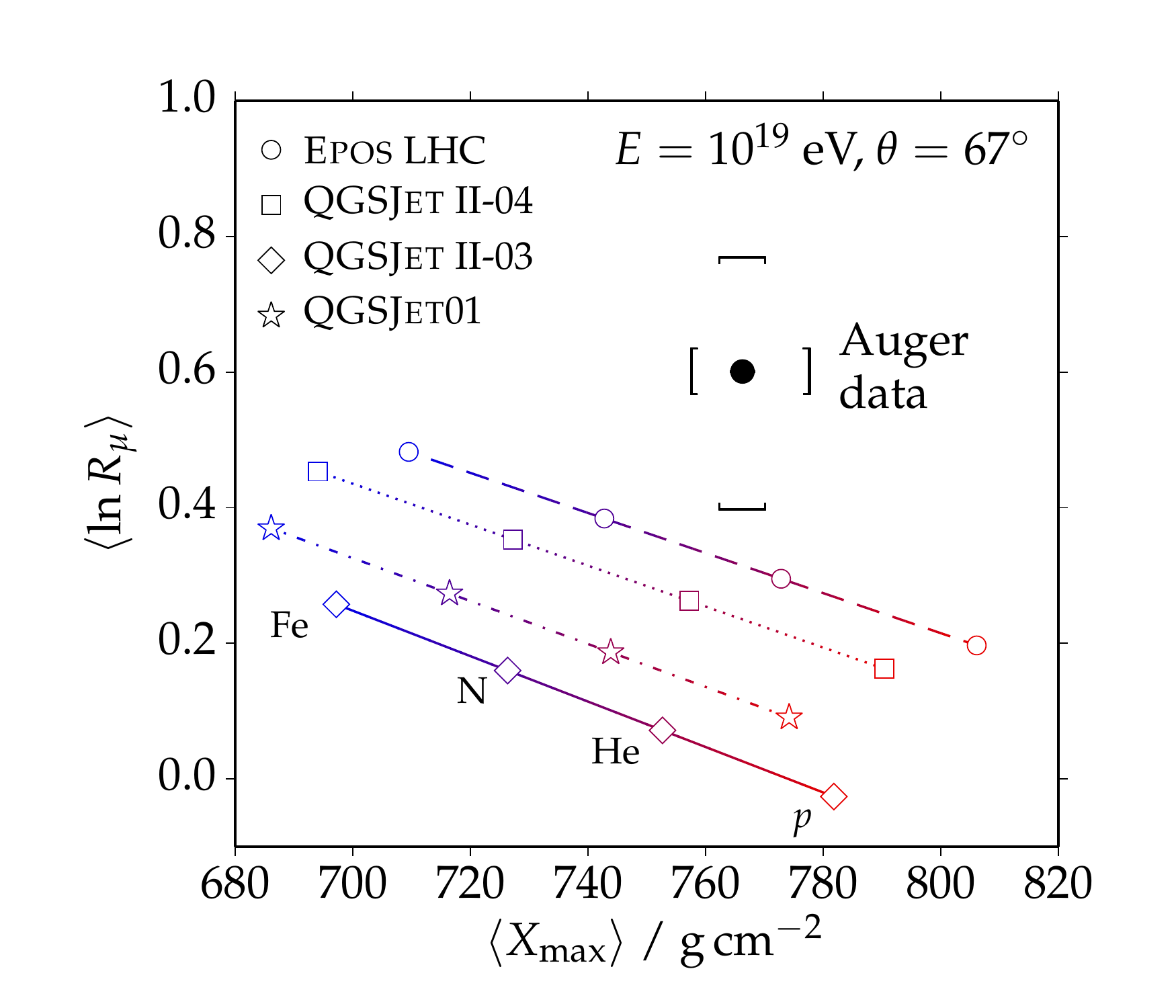}
	\vspace{-0.2cm}
	\caption{Results of the analysis of the muon content
		in highly inclined events~\cite{\AugerHASMuonPaper}.}
	\label{fig:has}
	\vspace{-0.4cm}
\end{figure*}

\section{Number of muons in highly inclined showers}
\label{sec:has}

An efficient way of measuring muons using the SD
without contamination from the electromagnetic component 
is by using the atmosphere as a shield. Above a certain 
atmospheric depth ($\sim 2000\;\text{g/cm}^2$), 
the electromagnetic component is strongly attenuated while
most muons still penetrate down to the ground. 
Such large atmospheric depths are reached by 
highly inclined events. In this analysis, events 
with zenith angle $\dg{62} < \theta < \dg{80}$ measured by both SD
and FD are selected and their muon content is reconstructed.
The full analysis description and results can be found
in Ref.~\cite{\AugerHASMuonPaper}. 

Because of the complicated dependencies of the muon
density ($\rho_\mu$) with the zenith angle ($\theta$)
and azimuthal angle ($\phi$) of an event and the 
axis distance of a station ($r$),
a template fit method was adopted 
to describe the distribution of muons on the ground
of measured events. The templates were built using Monte Carlo
simulations of proton primaries at \E{19} using \QGSJetOldLong~\cite{\QGSJetPaper} 
as the hadronic model. The measured SD signals were then fitted by using the 
templates in which the normalization parameter ($N_{19}$) was left free.
By applying the same fit to simulated events, the bias on $N_{19}$
was estimated. After correcting for the bias, an unbiased estimator
for the template normalization $R_\mu$ was obtained. For any given 
zenith angle, \nmu can be recovered from $R_\mu$. For example,
$R_\mu=1$ corresponds to $1.2\times 10^7$ muons 
at the ground level above 0.3 \GeV (muon energy threshold in the WCDs) 
for $\theta = \dg{70}$. 
  
The average $R_\mu$ over energy bins as a function of energy 
is shown in~\cref{fig:has} (left), where the energy was reconstructed
independently of $R_\mu$ from the FD measurements. 
The predictions for proton and iron primaries 
using two hadronic models are also shown for comparison. 
The systematic uncertainties, depicted with square brackets,
are dominated by the uncertainties on the energy scale. 

By comparing our measurements with predictions of simulations,
we can observe that, even considering the lowest values 
of $\langle R_\mu\rangle$ allowed by the systematic uncertainties,
the muon content of inclined events would imply a very
heavy composition interpretation. The compatibility of this 
interpretation with the composition scenario expected from the \xmax
measurements is tested in~\cref{fig:has} (right), where we
show $\langle \ln R_\mu \rangle$ versus \xmaxmean at \E{19}. 
One can conclude that, by assuming the \xmax composition,
the measured value of $R_\mu$ is significantly larger than
that expected from the simulation, regardless of the hadronic model. 
This fact shows another manifestation of the 
muon-deficit problem, which is complementary to the result
presented in~\cref{sec:topdown} in terms of the event zenith angle.

\begin{figure*}[!ht]
	\centering
	\includegraphics[width=0.9\textwidth]{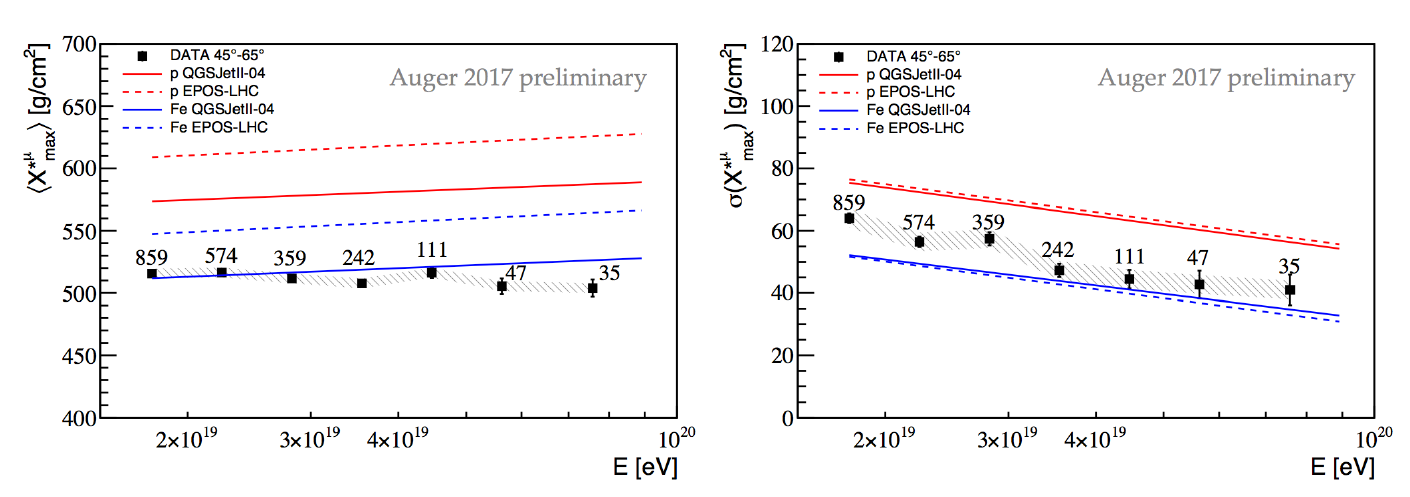}
	\vspace{-0.2cm}
	\caption{Results of the muon production depth
		analysis~\cite{AugerMPDICRC}.}
	\label{fig:mpd}
	\vspace{-0.4cm}	
\end{figure*}

\section{Muon production depth}
\label{sec:mpd}

The time structure as measured by the SD contains 
a lot of information about the EAS development. 
The analyses presented in the next three sections make
use of the time traces recorded by the WCDs to reconstruct EAS
properties. 

Since a muon propagates nearly linearly from its production
point down to the ground, its arrival time can be mapped into
their production depths by using simple geometrical considerations. 
In practice, this procedure requires SD stations with a large 
muon purity and a precise knowledge of the muon time delay. 
The former requirement is satisfied by selecting 
stations far from the shower axis in highly inclined showers,
while a model based on simulations is used to correct
for the muon time-delay. 

After reconstructing the muon production depth for each event,
the maximum of the profile (\xmumax) can be determined, the energy evolution
of its moments can be evaluated, and compared to predictions
by simulations. The \xmumax carries the information about the
depth of the first interaction, folded with the longitudinal development
of the hadronic component of the EAS which is responsible for the muon
production. It has been shown that \xmumax is very sensitive
to the properties of pion-air
interactions~\cite{Ostapchenko:2016bir,Pierog:2015ifw}.

A first version of the muon production depth analysis 
was published in Ref.~\cite{\AugerMPDPaper}, and recently,
an updated version has been released in Ref.~\cite{AugerMPDICRC}.
Among other improvements, the updated analysis extends 
the zenith angle and the lateral distance range of the selected
stations, increasing significantly the number of available events.
With the resulting larger statistical power, it is possible to derive not only
the \xmumaxmean, but also the \xmumaxsig. In~\cref{fig:mpd}, we show
both \xmumaxmean and \xmumaxsig as a function of energy in comparison
with the prediction from simulations of proton and iron primaries. 
Similarly to what can be seen in the results published in
Ref.~\cite{\AugerMPDPaper}, the composition interpretation of
the \xmumaxmean implies the presence of primaries as heavy as 
iron for \QGSJetLong~\cite{Ostapchenko:2004ss} 
and even heavier for \EposLHCLong. 
This interpretation is clearly inconsistent with that
obtained from the \xmax measurements, which means
that the hadronic models cannot properly describe 
the longitudinal development of muon production yet. 

\begin{figure*}[!ht]
	\centering
	\includegraphics[rviewport=0 0 0.93 1, width=0.355\textwidth]{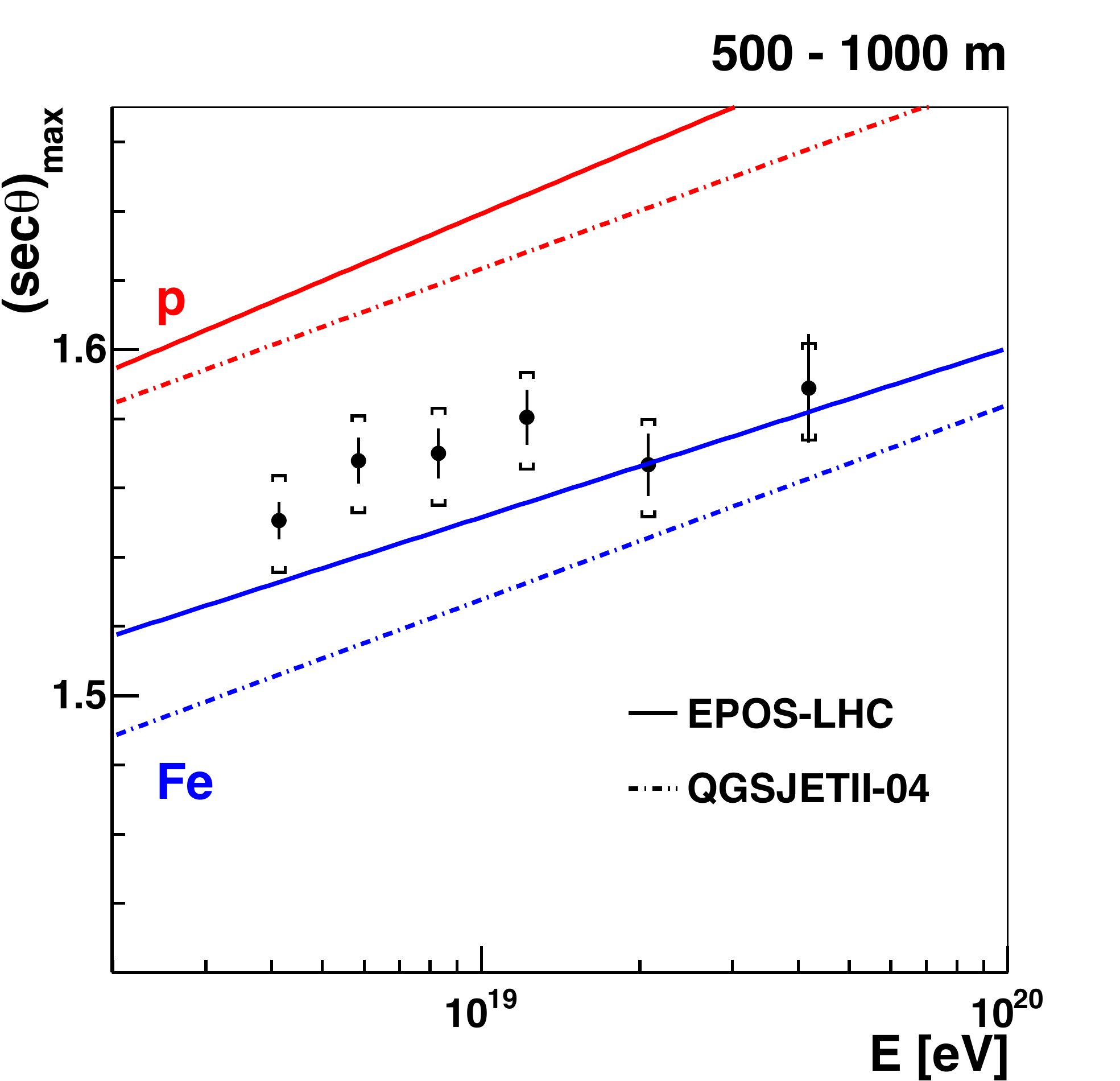}
\hspace{1cm}
	\includegraphics[rviewport=0 0 0.93 1, width=0.355\textwidth]{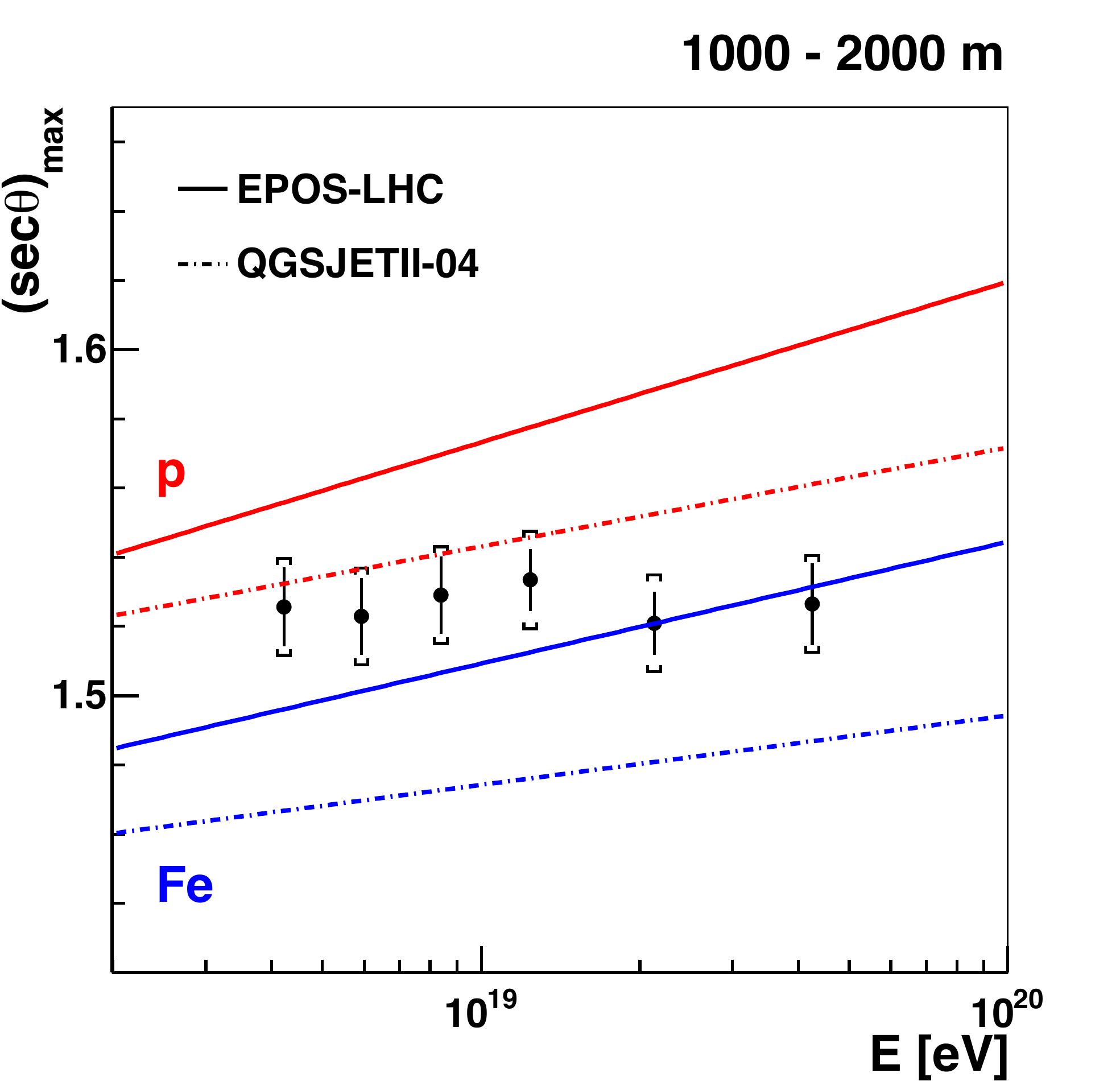}
	\vspace{-0.2cm}
	\caption{Results of the azimuthal
		asymmetry analysis of risetime~\cite{\AugerAsymmetryPaper}.}
	\label{fig:asym:moments}
	\vspace{-0.4cm}
\end{figure*}

\section{Azimuthal asymmetry of the risetime}
\label{sec:asym}

Both analyses presented in this and in the next section
are based on the risetime parameter (\rt).
It is defined as the time of increase 
from 10\% to 50\% of the integrated signal of each WCD station. 
Given that muons on average arrive earlier at the detector
than electrons, \rt turns out to be sensitive to the relationship
between these two components. This implies that, for example,
increasing the relative number of muons arriving 
in a given station would result in a faster increase in the WCD signal 
and thus a smaller \rt. 
Because of this, the \rt parameter is suitable for composition
inferences and tests of hadronic models.

Because \rt is a station quantity with a very complicated dependence
on parameters like shower zenith-angle, station azimuthal-angle 
and station lateral-distance, it is impractical to define one 
unique \rt-based parameter for each event. Instead, 
the two analyses presented in this paper follow approaches
in which these above mentioned dependencies are explored
in order to obtain information about the EAS development.
In this section, the azimuthal dependence of \rt will be explored.

It is observed that the average \rt as a function of the station
azimuthal angle ($\zeta$) presents a maximum at the point in which
the shower front reaches the ground the earliest, where the 
relative contribution of the electromagnetic 
component to the signal is the largest. This point is defined as 
$\zeta=0$. The shape of \rtmean as a function of $\zeta$ can be well
described by $\rtmean/r = a + b\cos\zeta+c\cos^2\zeta$, where the 
factor $r$ is included to account for the almost linear dependence
of \rtmean on the distance $r$ from the shower. The degree of asymmetry of
\rtmean can be quantified by the asymmetry factor defined as $b/(a+c)$.
For larger asymmetry factors, the peak of \rtmean at $\zeta=0$ is more
pronounced. 

In nearly vertical events, the asymmetry factor is expected to vanish
for symmetry reasons. 
The same behavior is expected for very inclined showers, in which
the shower front is dominated by muons that are very weakly attenuated
and consequently present very similar time structures 
for all azimuthal regions.
In between these two extremes, the asymmetry factor presents a maximum. 
It is observed that the zenith angle which corresponds 
to this maximum ($\theta_\text{max}$)
carries information about the relative amount of muons in EAS and consequently 
can be used as an observable sensitive to composition 
and hadronic interaction properties.
The full description of the analysis and results can be found in Ref.~\cite{\AugerAsymmetryPaper}.

In~\cref{fig:asym:moments}, we show the energy evolution of
$\sec\theta_\text{max}$ derived in energy bins. The predictions
from simulations are also shown for two hadronic interaction models
for proton and iron primaries. The left (right) plot shows the results
for the sub-dataset of stations with $500 < r/\text{m} < 1000$ 
($1000 < r/\text{m} < 2000$). The composition interpretation
of the $\sec\theta_\text{max}$ measurements 
imply, in general, the presence of heavier primaries on average,
when compared to the composition expected from the \xmax measurements
(see Ref.~\cite{\AugerAsymmetryPaper}).     
The difference is of order of $\Delta\ln A \approx 1$. 
The only case in which both observables 
are consistent is for \QGSJetLong at $1000 < r/\text{m} < 2000$.
The inconsistency with the \xmax composition and between the two 
different axis distance ranges, for \QGSJetLong, are indications
of a problem with the description of the \rt asymmetry by the models. 
Since the \rt is sensitive to the muon content of EAS,
these problems can be related to the muon deficit. 

\section{The Delta method}
\label{sec:delta}

\begin{figure*}[!h]
	\centering
	\includegraphics[rviewport=0.05 0 0.92 0.94, width=0.355\textwidth]{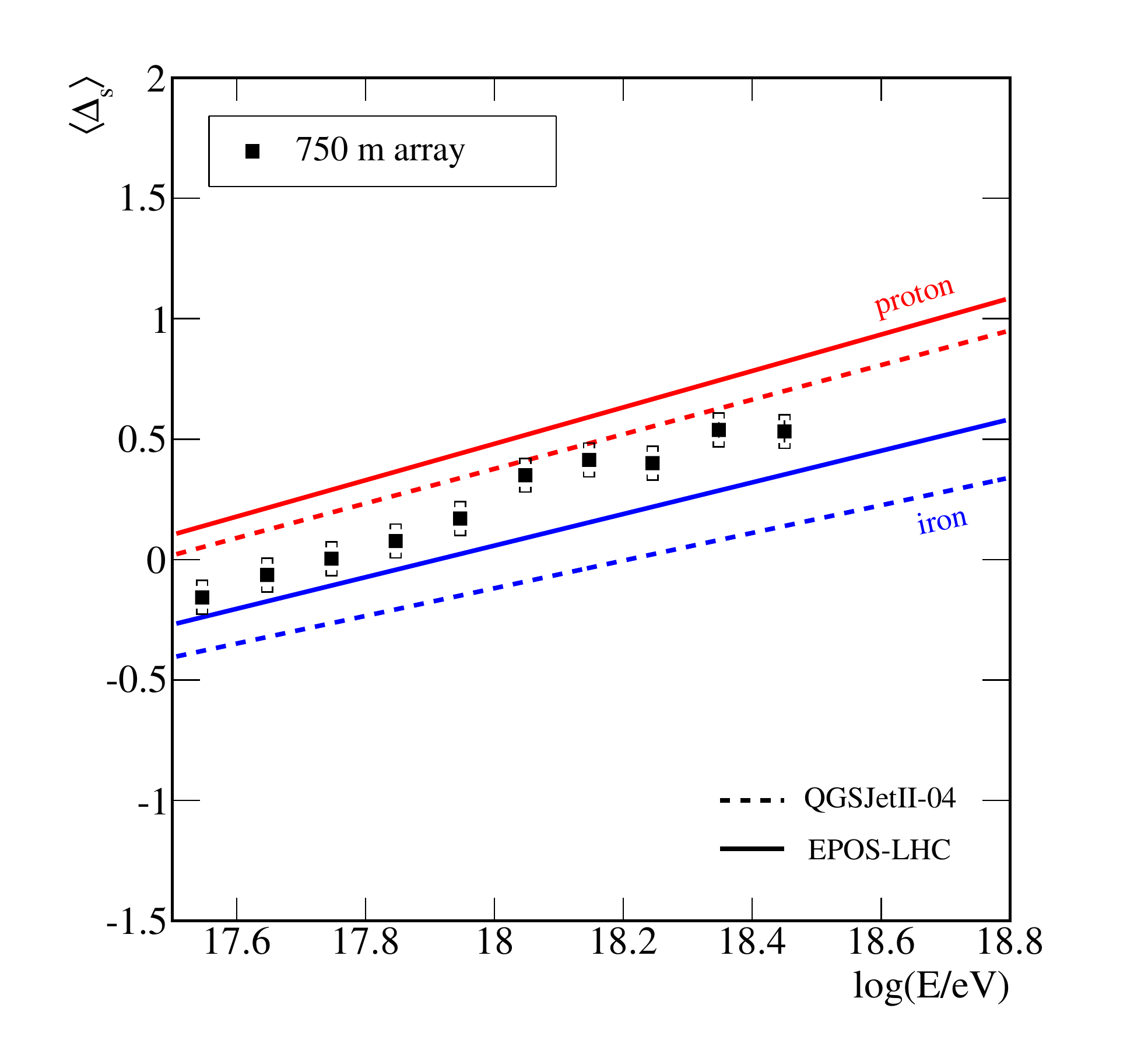}
	\hspace{1cm}
	\includegraphics[rviewport=0.05 0 0.92 0.94, width=0.355\textwidth]{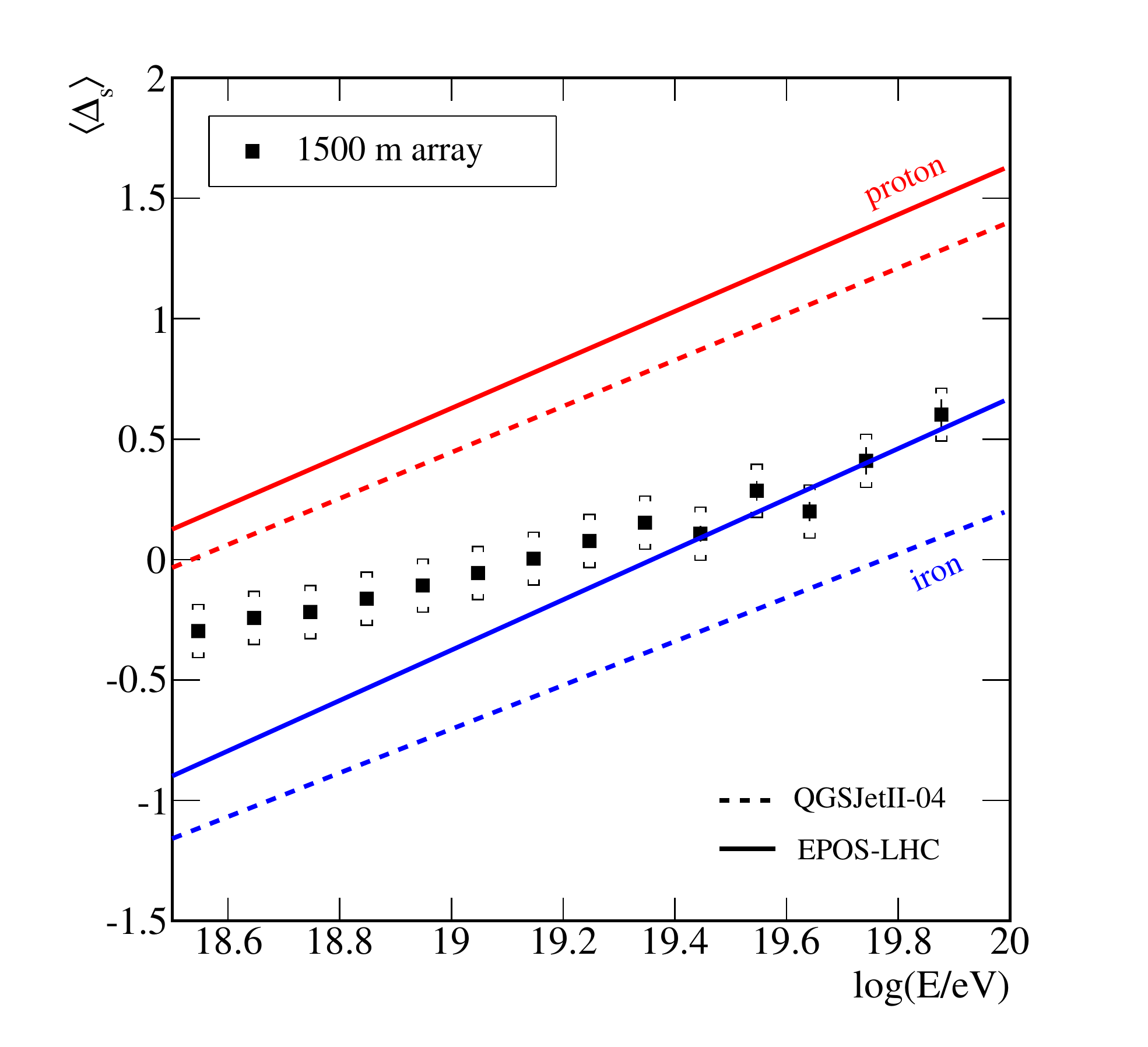}
	\vspace{-0.2cm}
	\caption{Results of the \dsmean method 
analysis~\cite{\AugerDeltaPaper}.}
	\label{fig:delta:moments}
	\vspace{-0.4cm}
\end{figure*}

The second analysis based on risetime to be presented in this
paper makes use of the dependence of \rt on the distance $r$ from 
the shower axis
to derive a new event parameter \ds. 
From the \rt measured for each SD station within an event,
the parameter 
$\Delta_i = (t_{1/2} - t_{1/2}^\text{bench})/\sigma_{\rt}$
is computed, where $t_{1/2}^\text{bench}$ is obtained from
a benchmark and $\sigma_{\rt}$ is the uncertainty on \rt. 
\ds is defined as the average of $\Delta_i$ over 
$N$ selected stations, $\ds = \sum_i \Delta_i/N$.

Two intermediate steps are important for the derivation of \ds:
the definition of the benchmarks and the estimation
of the \rt uncertainties. First, the benchmark is defined 
as the average behavior of \rt as a function of $r$ observed
in data for a given reference primary energy. By construction,
the \dsmean vanishes at the reference energy. The benchmark
curve was parametrized accounting for its dependence on
the zenith angle.
Secondly, the \rt uncertainties are determined through a detailed study 
of the subsets of SD twins stations, which
are pairs of stations located effectively at the same position,
and \emph{pairs}, which are stations within the same event at
different positions but at approximately the same distance from
the shower axis. 

A detailed parametrization of $\sigma_{\rt}$ 
was then performed by accounting for its dependences on $r$, $\theta$,
and the station signal $S$.  
The full description of the method and the results can be 
found in Ref.~\cite{\AugerDeltaPaper}. 
 
In~\cref{fig:delta:moments}, we show the \dsmean in energy bins as 
a function of the primary energy together with the predictions
by simulations with two hadronic models. The two plots show 
the results for the two SD arrays, with SD stations spaced
by 750 m on the left and 1500 m on the right. The energy
bins chosen for the definition of the benchmarks are
$17.7 < \lg(E/\text{eV})< 17.8 $ and $19.1 < \lg(E/\text{eV})< 19.2 $
for the 750 and 1500 m array, respectively. 
The composition interpretation from the comparison 
of the measured \ds with the predictions from simulations 
is not consistent with the one from the \xmax measurements 
(see Ref.~\cite{\AugerDeltaPaper}). 

Although strong similarities can be 
seen on the energy evolution of \lnamean, a systematic difference
of $\Delta \ln A \approx 1$ to $1.5$ is observed, where the composition
derived from \ds is heavier than the one from \xmax. These differences
on \lnamean are of the same order as the differences observed
in the \rt azimuthal asymmetry analysis (see~\cref{sec:asym}). 
Thus, the \dsmean method provides us with further evidence that
the hadronic models cannot properly describe the \rt parameter,
assuming that the uncertainties on the \xmax predictions are 
relatively small. Again, this can be related to the muon
deficit problem, since the \rt is sensitive 
to both electromagnetic and muonic components.  
 
It is worthwhile noting that, apart from 
testing the predictions of the hadronic models,
the \ds method can also be effectively used for 
composition inferences through a cross calibration
procedure with the \xmax measured by the FD
(see Ref.~\cite{\AugerDeltaPaper} for the composition
results). 

\section{Summary}
\label{sec:summary}

The experimental setup of the Pierre Auger 
Observatory offers great capability for 
testing hadronic interactions.
While the calorimetric energy and the longitudinal
development of the showers (\xmax) can be measured by the FD, 
the SD provides the lateral distribution of the
particles on the ground as well as the particle arrival
time distributions. By combining measurements of both
detectors, it is possible to reconstruct a number
of properties of EAS that can be used to test
the consistency of the EAS description
with the simulations with hadronic interaction models. 

We have presented here a selection of five analyses.
In~\cref{sec:topdown,sec:has}, the muon content
of EAS was measured with two different approaches and 
using datasets with complementary zenith angle ranges.
The muon deficit in simulations was observed in both cases.
An increase of at least 30\% on the number of muons
in simulation predictions is required to make
the muon content measurements consistent with the 
composition interpretation from the \xmax measurements.

In~\cref{sec:mpd}, the measured arrival time distributions
of the particles at the ground were used to access the
longitudinal development of the muonic component. 
The maximum of the muon production depth was reconstructed 
and, by comparing its moments to simulation predictions,
it is observed that the hadronic models cannot provide
a consistent description of the muon production profile and the 
\xmax. 

In~\cref{sec:asym,sec:delta}, two analyses based on the \rt 
parameter were presented: the \rt azimuthal asymmetry analysis and
the \ds method. The results from both analyses show again an
inconsistency in the descriptions of these observables when
using the \xmax as a reference. The discrepancy in this case is not
as large as in the case of the pure muonic observables (\nmu and \xmumax)
and it might be also related to the deficit of muons in simulations.

More precise constraints on the hadronic models will be possible 
in the future with AugerPrime, the upgrade of the Pierre Auger Observatory~\cite{Aab:2016vlz}. 
By using dedicated muon detectors
of two types (SSD and AMIGA), more information about 
the muonic component will be obtained and valuable
experimental input will be provided for the study of 
hadronic interactions.

\bibliography{lib}

\end{document}

%% file: TexAbbreviations.tex
\newcommand{\e}[1]{\ensuremath{\mbox{$10^{#1}$}}\xspace}
\newcommand{\E}[1]{\ensuremath{\mbox{$10^{#1}\;\eV$}}\xspace}

\newcommand{\dg}[1]{\ensuremath{\mbox{#1$^\circ$}}\xspace}

\newcommand{\eV}{\ensuremath{\mbox{e\kern-0.1em V}}\xspace}
\newcommand{\EeV}{\ensuremath{\mbox{Ee\kern-0.1em V}}\xspace}
\newcommand{\TeV}{\ensuremath{\mbox{Te\kern-0.1em V}}\xspace}
\newcommand{\PeV}{\ensuremath{\mbox{Pe\kern-0.1em V}}\xspace}
\newcommand{\GeV}{\ensuremath{\mbox{Ge\kern-0.1em V}}\xspace}
\newcommand{\MeV}{\ensuremath{\mbox{Me\kern-0.1em V}}\xspace}
\newcommand{\GeVc}{\ensuremath{\mbox{Ge\kern-0.1em V}\kern-0.1em/\kern-0.05em c}\xspace}
\newcommand{\GeVcc}{\ensuremath{\mbox{Ge\kern-0.1em V}\kern-0.1em/\kern-0.05em c^2}\xspace}
\newcommand{\AGeV}{\ensuremath{A\,\mbox{Ge\kern-0.1em V}}\xspace}
\newcommand{\AGeVc}{\ensuremath{A\,\mbox{Ge\kern-0.1em V}\kern-0.1em/\kern-0.05em c}\xspace}
\newcommand{\MeVc}{\ensuremath{\mbox{Me\kern-0.1em V}\kern-0.1em/\kern-0.05em c}\xspace}
\newcommand{\MeVcc}{\ensuremath{\mbox{Me\kern-0.1em V}\kern-0.1em/\kern-0.05em c^2}\xspace}
\newcommand{\T}{\ensuremath{\mbox{T}}\xspace}

\newcommand{\QGSJetLong}{{\scshape QGSJet\,II-04}\xspace}
\newcommand{\QGSJetOldLong}{{\scshape QGSJet\,II-03}\xspace}

\newcommand{\EposLHCLong}{{\scshape Epos\,LHC}\xspace}

\newcommand{\CernVM}{\textsc{Cern\-\kern-0.05emVM}\xspace}

%% file: ThisAbbreviations.tex
\newcommand{\nmu}{\ensuremath{\mbox{$N_\mu$}}\xspace}

\newcommand{\xmax}{\ensuremath{\mbox{$X_\text{max}$}}\xspace}
\newcommand{\xmaxmean}{\ensuremath{\mbox{$\langle X_\text{max}\rangle$}}\xspace}

\newcommand{\xmumaxsig}{\ensuremath{\mbox{$\sigma[X^{\mu}_\text{max}]$}}\xspace}
\newcommand{\xmumax}{\ensuremath{\mbox{$X^\mu_\text{max}$}}\xspace}
\newcommand{\xmumaxmean}{\ensuremath{\mbox{$\langle X^\mu_\text{max}\rangle$}}\xspace}

\newcommand{\lnamean}{\ensuremath{\mbox{$\langle \ln A\rangle$}}\xspace}
\newcommand{\rt}{\ensuremath{\mbox{$\text{t}_{1/2}$}}\xspace}
\newcommand{\rtmean}{\ensuremath{\mbox{$\langle\text{t}_{1/2}\rangle$}}\xspace}
\newcommand{\ds}{\ensuremath{\mbox{$\Delta_s$}}\xspace}
\newcommand{\dsmean}{\ensuremath{\mbox{$\langle\Delta_s\rangle$}}\xspace}
\newcommand{\size}{\ensuremath{\mbox{$S_{1000}$}}\xspace}


%% file: bib_aliases.tex
\def\T2KPaper{Abe:2011ks}

\def\AugerPaper{ThePierreAuger:2015rma}

\def\QGSJetPaper{Ostapchenko:2010vb}

\def\EposLHCPaper{Pierog:2013ria}

\def\AugerHASMuonPaper{Aab:2014pza}
\def\AugerMPDPaper{Aab:2014dua}

\def\AugerTopDownPaper{Aab:2016hkv}
\def\AugerXmaxPRDPaper{Aab:2014kda}
\def\AugerXmaxICRC2017Paper{Bellido:2017cgf}
\def\AugerAsymmetryPaper{Aab:2016enk}
\def\AugerDeltaPaper{Aab:2017cgk}

\def\HiResMIAMuonPaper{AbuZayyad:1999xa}
\def\IceTopMuonPaper{Dembinski:2017zkb}

\def\SugarMuonPaper{Bellido:2018toz}

\def\APP16{Prado:2016akv}
\def\APP17{Prado:2017ijs}